\documentstyle[12pt,aasms4]{article}
\begin{document}

\title{Absolute Proper Motions to B $\sim 22.5$:
V. Detection of Sagittarius Dwarf Spheroidal Debris in the Direction
of the Galactic Anticenter}

\author{Dana I. Dinescu\altaffilmark{1,3}, Steven R. Majewski\altaffilmark{2,4}, Terrence M. Girard\altaffilmark{1}, R\'{e}ne A. M\'{e}ndez\altaffilmark{5}, 
Allan Sandage\altaffilmark{6},
Michael H. Siegel\altaffilmark{2,7}, 
William E. Kunkel\altaffilmark{8}, 
John P. Subasavage\altaffilmark{2},
and Jamie Ostheimer\altaffilmark{2}}

\altaffiltext{1}{Astronomy Department, Yale University, New Haven, CT 06520-8101 (dana@astro.yale.edu, girard@astro.yale.edu)}
\altaffiltext{2}{Department of Astronomy, University of Virginia, Charlottesville, VA 22903-0818 (srm4n@virginia.edu)}
\altaffiltext{3}{Astronomical Institute of The Romanian Academy, Cutitul de Argint 5, RO-75212 Bucharest 28, Romania}
\altaffiltext{4}{David and Lucile Packard Foundation Fellow, Cottrell Scholar of the Research Corporation} 
\altaffiltext{5}{European Southern Observatory, Casilla 19001, Santiago 19, Chile}
\altaffiltext{6}{The Observatories of the Carnegie Institution of Washington,
813 Santa Barbara Street, Pasadena, CA 91101}
\altaffiltext{7}{Space Telescope Science Institute, 3700 San Martin Drive, Baltimore, MD 21218}
\altaffiltext{8}{Las Campanas Observatory, Casilla 601, La Serena, Chile}

\begin{abstract}

We have detected a population of predominantly blue
$(B-V \le 1.1)$ stars in the direction $l = 167\arcdeg, b = -35\arcdeg$
(Kapteyn Selected Area 71) that cannot be accounted for by 
standard starcount models. Down to $V \sim 20$, the colors and 
magnitudes of these stars are similar to those of the southern overdensity 
detected by the Sloan Digital Sky Survey at $l = 167\arcdeg, b = -54\arcdeg$,
and identified as stripped material from the Sagittarius dwarf 
spheroidal galaxy.
We present absolute proper motions of the stars in SA 71, and 
we find that the excess blue stars represent a distinct, 
kinematically cooler component than the Galactic field, and 
in reasonable agreement with 
predictions of Sgr disruption models.
The density of the excess SA 71 stars 
at $V \sim 18.8$ and $B-V \le 1.1$ is within a factor of two of the
density of the SDSS-south Sgr stripped material, and of that
predicted by the Helmi \& White disruption model.
Three additional anticenter fields 
(SA 29, 45 and 118) show very good agreement with 
standard starcount models.

\end{abstract}


\section{Introduction}

The Sagittarius dwarf galaxy (Sgr) 
(Ibata, Gilmore \& Irwin 1994) has drawn considerable interest 
as a ``living'' example of the accretion process supposedly
responsible for the 
creation of the outer Galactic halo. 
Most early Sgr surveys focused  
within $10$-$15\arcdeg$ of its 
core (e.g., Ibata {\it et al.} 1997, hereafter I97, Mateo {\it et al.} 1998). 
More recently, 
Sgr-associated material has been identified farther from its main body
(e.g., Majewski {\it et al.} 1999, hereafter M99), and
it is becoming apparent that Sgr tidal debris may
envelop the whole Galaxy roughly along the satellite's orbit.
Results from the Sloan Digital Sky Survey
(SDSS) in two equatorial strips 
(Yanny {\it et al.} 2000, hereafter Y00;
Ivezic {\it et al.} 2000, hereafter I00) find 
RR Lyrae stars, blue horizontal branch (BHB) stars, and/or
blue stragglers (BS) in excess
of the Galactic population that are
most plausibly explained as Sgr material 
(see Ibata {\it et al.} 2001a).
Newberg {\it et al.} (2002, hereafter N02) confirm the previous SDSS 
Sgr detections with color-magnitude diagrams (CMDs) showing main sequence turnoff 
colors (MSTO) consistent with those of Sgr.
Similarly, Mart\'{i}nez-Delgado {\it et al.} (2001a,b;
collectively referred to as M01) report Sgr detections associated with 
MSTO stars in several regions along the Sgr orbital path. Vivas {\it et al.} (2001) 
confirm the I00 detection of 
RR Lyrae stars in a $\sim 50$ kpc distant clump 
at positive Galactic latitudes, while 
Dohm-Palmer {\it et al.} (2001) and Kundu {\it et al.} (2002; K02 hereafter)
report giant star concentrations in the inner halo having radial velocities (RVs)
and distances in good agreement with wrapped tidal arms predicted from
Sgr disruption models. Ibata {\it et al.} (2001b) also show that 
the carbon-star distribution from the 
Totten \& Irwin (1998, hereafter TI98) all-sky survey
is highly correlated with the orbital path of Sgr.

Here we present absolute proper motions and $B-V$ 
color  distributions in Kapteyn Selected Area (SA) 71, 
a Galactic anticenter field located close to the 
orbital path of Sgr
($ l = 167 \arcdeg, b = -35 \arcdeg$). We find 
a population of predominantly blue stars ($ B-V \le 1.1$)
in excess of that
predicted by standard Galactic starcount models. 
The proper motions of the excess blue stars show a distinct, kinematically 
cooler population than the Galactic field. 
Assuming that the excess blue stars are horizontal branch (HB) stars,
we estimate their distance to be $\sim 30$ kpc, a value consistent with
Sgr disruption model predictions (e.g., Johnston 1998, hereafter J98).
We also present $B-V$ color distributions for three other anticenter fields 
(SA 29, 45 and 118). These allow us to refine some of the
parameters of the starcount model. These three fields are consistent with each other and agree with the global model predictions. 
This work is a continuation of the 
photometric and proper motion study of SA 57 by Majewski (1992) as well as 
the Mount Wilson Halo Mapping Project (Sandage 1997).

\section{The Data}

Star catalogs are derived from typically six Mayall 4m prime focus photographic plates 
(plate scale $18\farcs60$ mm$^{-1}$) per field: 
two each in the $J$ (IIIa-J+GG385), $F$ (IIIa-F+GG495) and $R$
(127-04 + RG610) passbands. In this paper we present
only the $JF$ photometry data which we transform to the Johnson $BV$ system
(see Majewski 1992). The plates were digitized over a 
$40\farcm3 \times 40\farcm3$ area
with the University of Virginia's Perkin-Elmer microdensitometer. 
The detection of objects, derivation of various image
parameters and object classification were performed 
using the FOCAS software package (Valdes 1982, 1993).

Calibration to standard Johnson $BV$ magnitudes used
Mount Wilson 100-inch  (MW) photoelectric
photometry (Sandage 1997, 2000) supplemented with:
CCD data for SA 71 and 118 obtained with the Swope 1m telescope 
at Las Campanas Observatory in 2001 and in 1993 respectively, 
Massey's (1998) SA 45 data, 
and, for SA 29, 
CCD photometry from the Mayall 4m and Mosaic camera.
From comparisons between the
calibrated photographic and standard photoelectric/CCD
magnitudes in the Johnson system,
we find uncertainties in 
the calibrated magnitudes between 
0.05 and 0.12, and in $B-V$ 
between 0.07 and 0.15, 
for the relevant magnitude range analyzed here.

Object classification relied on the resolution 
classifier described by Valdes (1982). Inspection of images and derived 
image parameters for different classes of objects shows 
reliable classification to  $B \sim 21.2$
($V \sim 20$). All fields are complete in terms of
object detection to $B \sim 22.5$, except
SA 118, which is complete to $B \sim 21.5$.
These ``completeness'' limits are 1.5 magnitudes above the plate limit.
For the deeper plates, stars brighter than $V = 17$ show varying
degrees of saturation, so 
we analyze only stars within $ 17 \le V \le 20$.
A detailed description of the calibration and object 
classification will be presented elsewhere. 

SA71 proper motions were determined from 
measurement of thirteen photographic plates spanning
a 22-year baseline: eleven Mayall 4m plates 
(of which six are the plates used for the 
photometry), and two modern Du Pont 2.5m plates with two exposures 
(short and long) each.
The emulsion+filter combination match the blue, visual and red passbands. 
The seeing varies between 1 to 3 arcsec, with the 
modern plates typically having  better images.
The photometric catalogs served as input lists for 
astrometric scans with the Yale PDS microdensitometer.
General procedures for the astrometric reduction are described elsewhere
(e.g., Girard {\it et al.} 1998, Dinescu {\it et al.} 2000).
We have made use of the double exposures on the Du Pont plates 
to model magnitude-dependent systematics, and have corrected
for differential color refraction for the appropriate photographic emulsion
and hour angle following the prescription in Dinescu {\it et al.} (1996).
Inspection of galaxy proper motions 
shows no detectable trends with colors and magnitudes.
We obtain a proper-motion uncertainty of 
0.8 mas yr$^{-1}$ per star for $V \le 19.5$, and $\sim 1$ mas yr$^{-1}$ for $V < 20.5$.
The correction to absolute proper motion is based on $\sim 500$
galaxies, but has an uncertainty of 0.2 mas yr$^{-1}$
due to poorer centering accuracy of the galaxy images.

\section{Results: Color Distributions and Proper Motions}

Stellar color distributions are shown in Figure 1.
In each panel, the total counts for the data (left numbers in 
parentheses)  and the starcount model (right numbers) are given.
The starcount and kinematical model used here 
includes representations of the thin and thick disks and the halo.
Details of the model can be found in M\'{e}ndez \& van Altena (1996, MvA96) and
M\'{e}ndez \& Guzm\'{a}n (1998).
The ``best-fit'' model parameters were determined by eye
from comparisons of all
four fields in three $V$-magnitude bins: 17-18, 18-19, and 19-20.
We adopt a maximum thin disk scale-height of 300 pc 
to represent late-type main sequence stars and white dwarfs.
The best-fit thick disk local normalization is $6\%$ of the thin disk with 
a 1000 pc scaleheight, with a 3.0 kpc  
scalelength for both disks.
The halo is an oblate ($c/a =0.8$) de Vaucouleurs law 
spheroid of local normalization $0.13\%$ to the thin disk.  
The reddening for a given distance is derived
from the extinction
model presented in M\'{e}ndez \& van Altena (1998), 
with infinite distance 
values from the Schlegel, Finkbeiner \& Davis (1998, hereafter SFD) maps
for the center of each SA field
(SA 29, $E_{B-V} = 0.012$; SA 45, $E_{B-V} = 0.054$; SA 71,
$E_{B-V} = 0.219$; and SA 118, $E_{B-V} = 0.025$).
The model color counts were convolved with the estimated 
photometric $B-V$ errors.
Field SA 45 partly contains the galaxy M33 in the east part of the field. 
Therefore, we have analyzed
the westernmost $18\farcm5 \times 40 \arcmin$ region of this field, which
shows no M33-contamination in the CMD.  
For field SA71, which has the largest reddening,
we have also determined star counts after dereddening star-by-star 
according to the
SFD maps. The comparison with model counts yields similar results to
when a single SA71 reddening value is applied.

The SA 29, 45 and 118 data agree well with model
predictions in all three magnitude bins (slight 
residual systematic errors may still exist in the photographic colors).
However, SA 71 shows a clear excess of blue stars
for $18 < V < 20$ compared to model predictions.
This excess occurs for $B-V \le 1.1$, and 
is most prominent in the $18 < V < 19$ bin.

In Figure 2, left panels, we show the vector point diagram (VPD) for 
$17 < V < 20$ stars in two color groups: 
$B-V \le 1.1$ (top panels) and 
$B-V > 1.1$ (bottom panels). 
The blue stars show a clump with dispersion 
$\sim 2$ mas yr$^{-1}$ and absolute mean proper motion
$\mu_{l}$ cos $b = 1.4 \pm 0.3$ and $\mu_b = -0.6 \pm 0.3$ mas yr$^{-1}$
superimposed on a field population with wider dispersion.
The absolute proper-motion uncertainty is given by 
the proper-motion measuring error for galaxies and
the proper-motion dispersion of the excess blue star clump. 
The red stars show a wide proper motion distribution typical of the Galactic
field. Intuitively, blue field stars have a lower dispersion than 
red ones, because red stars are predominantly (see Figure 1) nearby disk 
K-M dwarfs, which have larger proper motions, while blue stars are 
primarily more distant thick disk 
and halo stars.  However, the adjoining plots of 
$\mu_l~cos~b $ (middle panels) and $\mu_b$ (right panels) distributions
show more complex character. The continuous
lines represent the proper motion distributions predicted from the
MvA96 Galactic model, which evidently agree
reasonably well with the data for the red stars, whereas the blue sample
shows an excess population with a lower dispersion than that of the field.
We have also obtained preliminary proper motions in SA 29, 
a field located symetrically to SA 71 across the Galactic plane.
The SA29 proper-motion distributions do not show the excess blue star 
population seen in SA 71.

\section{What is the Excess Population?}

In Figure 3 we show in Aitoff projection 
the location of our SA fields, most 
known detections (filled symbols) and
non-detections (open symbols) of Sgr material, and the orbit of Sgr as
determined from the ground-based proper-motion measurement of I97
(continuous line).
SA 71 is located close to Sgr's orbit, and to a clump of five 
carbon stars found in the surveys of TI98 and Green {\it et al.} (1994).
These  five carbon stars belong to 
two groups: 1) three stars with 
$R$ magnitudes 15.1, 15.5, and 16.4 and
heliocentric RVs of -133, -140, and -140 km s$^{-1}$ 
respectively (highlighted crosses in Fig. 3); 
2) two stars with $R = 12.6$ and 13.0 and RVs
of 112 and 43 km s$^{-1}$ respectively. 
Except for the faintest carbon star, which is of CH type, 
all are N type; according to TI98,
the N type can be assigned an absolute magnitude
$M_{R} = -3.5$. Using the SFD reddening values
for each location of the carbon stars,
we obtain that the fainter group resides at distances 
35-44 kpc from the Sun. 
>From the SDSS-south survey (Y00, N02), which is close to SA 71 (Fig. 3), 
the distance to the candidate Sgr HB stars is $ \sim 28$ kpc.
The distances of the carbon stars and of the HB stars in 
SDSS-south match the distances predicted for tidal debris from Sgr at this
Galactic location (see e.g., Helmi \& White 2001, hereafter HW01; M01). 
Moreover, the 
large, negative RVs
of three of the carbon stars (the highlighted crosses in Fig. 3)
also match the model predictions for 
a stream from Sgr that consists of material lost
 at the previous pericenter passage
(Sgr is currently at pericenter, see e.g, HW01, 
Ibata {\it et al.} 2001b).

Considering the proximity of SA 71 to directions where
Sgr material has been detected,  we can plausibly associate the 
excess blue stars in SA 71 with Sgr material.
For $18 < V < 19$
the excess stars peak at  $B-V \sim 1.0$ ($[B-V]_0 =  0.78$)
a color consistent with Sgr's HB/red clump 
(Bellazzini, Ferraro \& Buonanno 1999). 
>From inspection of the SA 71 CMD, we adopt $V \sim 18.8$ as the 
magnitude of the candidate Sgr HB, corresponding to 
a heliocentric distance of 29-32 kpc (the distance range 
derives from the absolute magnitude spread for 
RR Lyrae stars of metallicities relevant to Sgr).
The estimated distance to the candidate Sgr HB is in agreement
with those of the carbon stars, the HB stars found in
SDSS-south, and model predictions (HW01, M01).
The excess population with $19 < V < 20$ 
is bluer than that with $18 < V <19$, and 
resembles the BS population found in Y00 and N02.

From kinematics of tidal stream material calculated with the 
semianalytical disruption model developed by J98 for Sgr and 
the measured orbit by I97, we find the nearest 
model Sgr stripped material to SA71 is at
$l = 174\arcdeg, b = -37\arcdeg$. This material belongs to a trailing stream
that emerged during the previous perigalacticon passage (0.75 Gyr ago). 
The debris is located 26 kpc from the Sun, 
and has $\mu_l~cos~b = 1.47$ mas yr$^{-1}$,
$\mu_b = -1.67$ mas yr$^{-1}$, and $V_{rad} = -134$ km s$^{-1}$. 
The vertical lines 
in the proper-motion distributions of Fig. 2 represent these predicted
proper motions (all are presented in the heliocentric rest frame).
To understand better the debris motion,
we subtract the reflex solar motion from our 
SA 71 proper-motion measurement, at a distance of
26 kpc. We obtain
$(\mu_l~cos~b)' = -0.47 \pm 0.30 $ mas yr$^{-1}$, and $\mu_b' = -0.35 \pm 0.30
$ mas yr$^{-1}$. 
For the model predictions we obtain 
$(\mu_l~cos~b)' = -0.43 $ mas yr$^{-1}$, and $\mu_b' = -1.55$ mas yr$^{-1}$.
While there is good agreement in $\mu_l~cos~b$, the 
predicted value of $\mu_b$ is larger in absolute
value than that measured.
This discrepancy may be due to uncertainties
in the model predictions stemming from the 
0.7 mas yr$^{-1}$ uncertainty in the
proper motion of Sgr (I97), and/or it may be due to 
undetected systematics in our proper-motion measurement. 
A full exploration of models with varying initial conditions 
may help understand this
discrepancy, but is beyond the purpose of this letter.
Given the almost polar orbit of Sgr (the latitude of the pole is
$b = 13\arcdeg$, Ibata {\it et al.} 2001b), 
one might want to compare the direction of our measured velocity vector 
to that estimated from orbit-shape considerations only.
However, we caution that, given the small size of the measured proper motion 
compared to its uncertainty, 
the direction is rather poorly constrained.
A small-size latitudinal motion 
is quite conceivable because debris in this particular stream reaches its
turning point below the Galactic plane at a Galactic location close to
that of SA 71 (see e.g., Fig. 3 in Ibata {\it et al.} 2001b).

Assuming a 1.7 mas yr$^{-1}$ lower limit
for the proper-motion dispersion 
(as given by the $\mu_b$ distribution of the excess blue stars), 
and an 0.8 mas yr$^{-1}$ measurement uncertainty, 
we obtain a 213 km s$^{-1}$ intrinsic velocity dispersion of the clump of blue stars
at 30 kpc. This dispersion is too large for a coherent structure to last for
0.75 Gyr. 
It is more plausible that we sample material lost in several
previous passages that is located at various distances, and thus has
a wide range in kinematics. 
The existence of multiple, overlapping Sgr streams is suggested by 
M99, Dinescu et al. (2000), Dohm-Palmer et al. (2001), and K02.
An alternative or additional explanation is that Galactic thick disk
stars contaminate our blue sample and thereby inflate the estimated 
dispersion (see Fig. 2).
As shown by K02, radial velocities can help disentangle 
streams left by previous Sgr passages,
(if this interpretation is correct) and also help separate 
field stars from  Sgr candidates.

The largest excess of blue stars is at $18 < V < 19$.
The difference between the data and the starcount model
for stars with $B-V \le 1.1$ is $47 \pm 15$, yielding a density of 
$0.029 \pm 0.009$ star arcmin$^{-2}$.
For the SDSS-south detection, N02 estimate a density three times
lower than that of SDSS-north. N02 find 500 red clump stars in the
SDSS-north along an orbital path of $4.4\arcdeg$. For our $40\farcm3$ field,
the SDSS-south 
density implies 25 red clump stars, or
$\sim 2$ times lower than our determination.
HW01 predict a rather high density of Sgr material at
$l = 179\arcdeg, b = -35\arcdeg$, a location close to SA 71. They 
derive a density relative to the main body of Sgr
of $\sim 0.005$ number-counts arcdeg$^{-2}$.
If the main body of Sgr
contains 34,000 red HB stars (Ibata, Gilmore and Irwin [1995] find 17,000 
red clump stars within the
half-mass radius of the main body of Sgr), then we should find 77 red HB
stars in our SA71 
area, according to the HW01 
model. This number is also within a factor of 2 of
our excess counts.

In the future, we plan to supplement this analysis with deeper, more accurate 
photometry, radial velocities and with detailed modeling of the dynamical
evolution of Sgr.

We thank M. Catelan for providing us HB models, 
M. Irwin for making available 
the carbon star data, and A. Helmi and K. Johnston for 
discussions concerning the Sgr disruption models.  
SRM thanks the Carnegie Observatories 
for access to Las Campanas facilities.
This research was supported by the NSF under grant AST 97-02521.

\newpage

\begin{figure}
\caption{$B-V$ color distributions for SA 29, 45, 71 and 118 compared to model 
predictions. 
The total model distribution is represented with 
the thick line.  The star counts of each Galactic component 
are also shown with thin lines. The thin disk dominates the star counts, with
smaller contributions from the thick disk and halo.
The numbers in each parenthesis represent the data/model counts.
}
\end{figure}

\begin{figure}
\caption{VPD (left panels) of stars between $V = 17 - 20$, 
and separated into two groups: blue ($B-V \le 1.1$; top panels) 
and red ($B-V > 1.1$; bottom panels).
The middle and right panels show the proper-motion distributions
in $\mu_l~cos~b$ and $\mu_b$ respectively
compared to model predictions. 
The vertical lines represent the 
predicted proper motion of a trailing tidal stream as calculated
with the J98 disruption model (see text).}
\end{figure}

\begin{figure}
\caption{Aitoff projection of current  detections (filled symbols) and 
associated non-detections (open symbols) of Sgr material. The projected orbit
is represented with a continuous line. The dotted line
represents the Galactic plane. Symbols are explained in the legend.}
\end{figure}

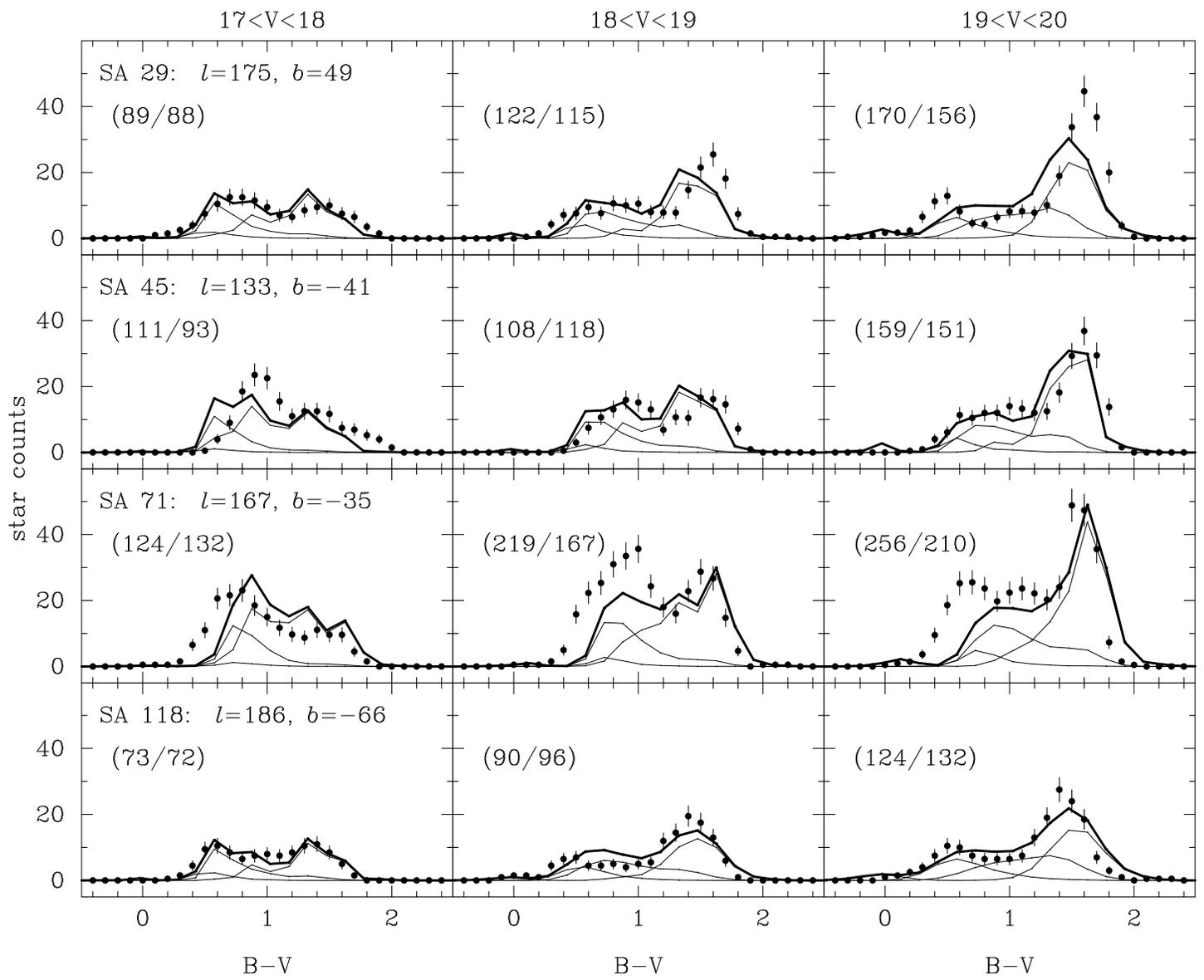
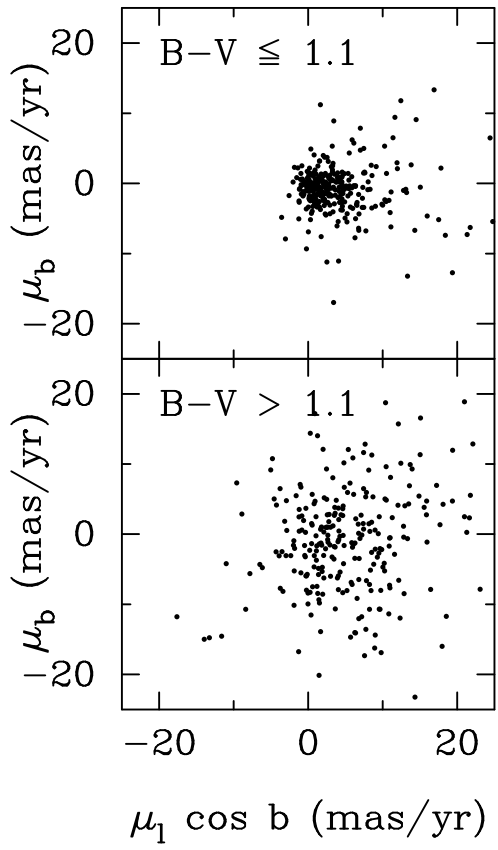
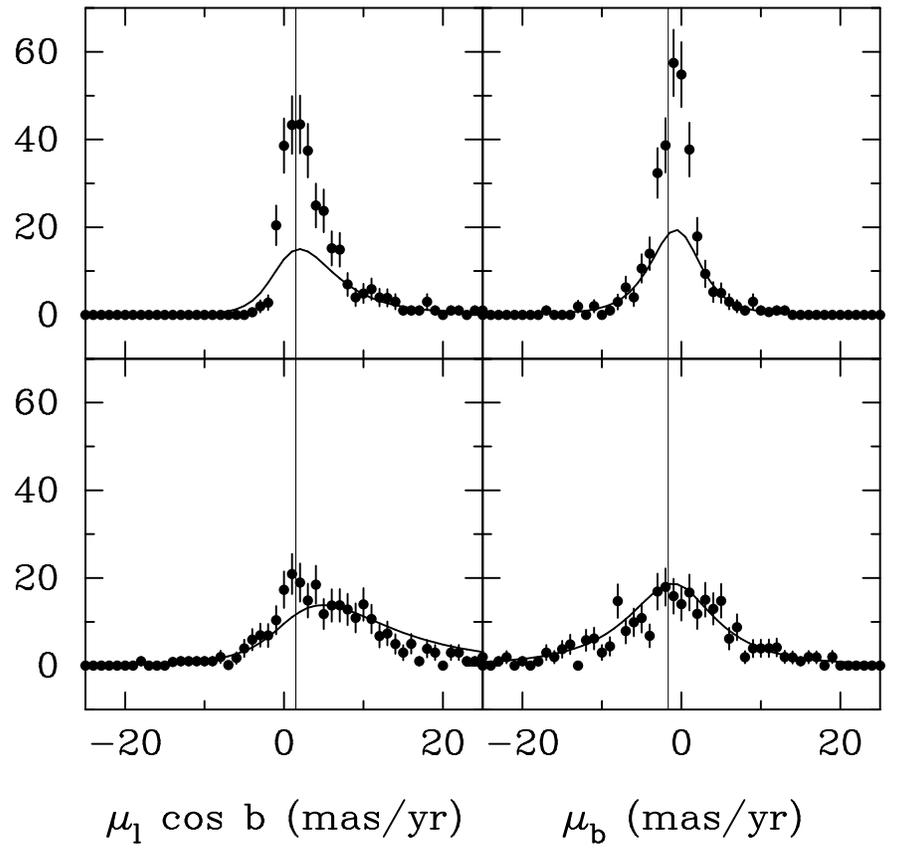
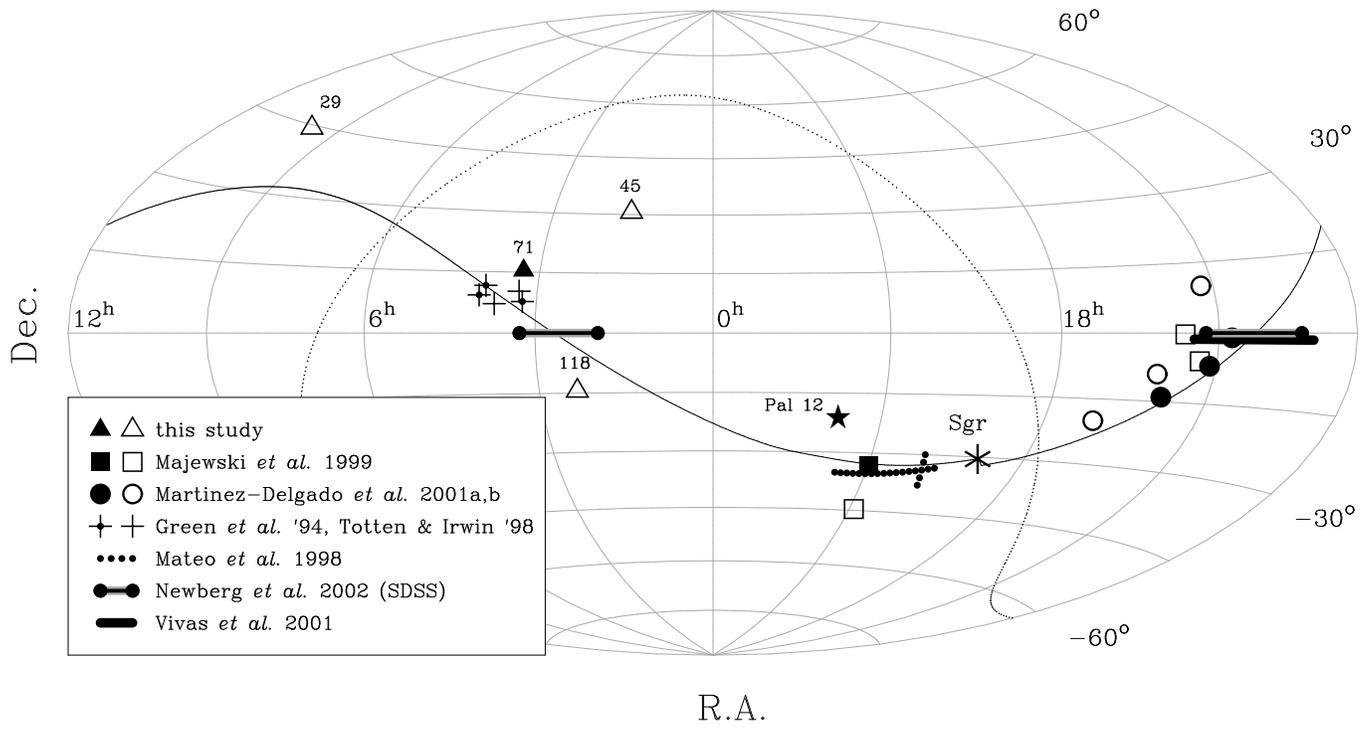
\end{document}